\documentclass[aps, prd, amsmath, floats, floatfix, twocolumn, nofootinbib, superscriptaddress]{revtex4-1}
\usepackage{graphicx}
\usepackage{color}
\usepackage{latexsym}
\usepackage{gensymb}
\usepackage[colorlinks]{hyperref}

\newcommand\be{\begin{equation}}
\newcommand\ba{\begin{eqnarray}}
\newcommand\ee{\end{equation}}
\newcommand\ea{\end{eqnarray}}

\begin{document}
\title{Iron Line Spectroscopy of Black Holes in Vector-Tensor Galileons Modified Gravity}

\author{Jinye Yang}
\affiliation{Center for Field Theory and Particle Physics and Department of Physics, Fudan University, 200438 Shanghai, China}
\author{Dimitry Ayzenberg}
\affiliation{Center for Field Theory and Particle Physics and Department of Physics, Fudan University, 200438 Shanghai, China}
\author{Cosimo Bambi}
\email[Corresponding author: ]{bambi@fudan.edu.cn}
\affiliation{Center for Field Theory and Particle Physics and Department of Physics, Fudan University, 200438 Shanghai, China}
\affiliation{Theoretical Astrophysics, Eberhard-Karls Universit{\"a}t T{\"u}bingen, 72076 T{\"u}bingen, Germany}

\date{\today}

\begin{abstract} 

Recently, a rotating black hole solution was found in Vector-Tensor Galileons modified gravity that has some significant differences from the Kerr black hole solution. We study the iron line shape that is part of the reflection spectrum of accretion disks around black holes in this new black hole solution. We simulate and compare the iron lines of this solution with those of the Kerr solution to see if the technique of iron line spectroscopy can be used as a tool to test Vector-Tensor Galileons modified gravity. Our analysis shows that current X-ray facilities can, in principle, be used to test and place constraints on Vector-Tensor Galileons modified gravity in the strong coupling and ultraspinning regimes. 

\end{abstract}

\maketitle

\section{Introduction}

According to the no-hair theorem, general relativistic black holes (BH) are completely described by the Kerr metric, which only depends on the mass and spin angular momentum. In addition, the no-hair theorem states that the Kerr metric is the only stationary, axisymmetric, asymptotically flat vacuum solution of the Einstein equations that has an event horizon and admits no external closed timelike curves~\cite{1967PhRv..164.1776I, 1968CMaPh...8..245I, 1971PhRvL..26..331C, 1972CMaPh..25..152H, 1975PhRvL..34..905R, 1973blho.conf...57C}. While modified gravity theories do not, in general, obey the no-hair theorem, many do admit the Kerr metric as a solution~\cite{Psaltis:2007cw}, and thus tests of the Kerr metric do not necessarily rule out all modified gravity theories. However, there are modified theories of gravity that do not have the Kerr metric as a solution, and in these cases any tests of the Kerr metric can also test these theories.

One such modified gravity theory that does not contain the Kerr metric as a solution is Vector-Tensor Galileons (VTG) modified gravity~\cite{Tasinato:2014eka, Gripaios:2004ms, Heisenberg:2014rta, Filippini:2017kov}. VTG gravity modifies General Relativity by including additional vector degrees of freedom that can be associated with dark matter or dark energy. Recently, an exact analytic rotating BH solution with regular horizons was found in~\cite{Filippini:2017kov}. The modifications to Kerr in this solution are parametrized by a dimensionless parameter $\beta$ and a charge $Q$ that is a charge associated with some dark force rather than the standard electromagnetic charge. When both $\beta$ and $Q$ vanish, the Kerr metric is recovered. A particularly interesting feature of this VTG BH solution is that for values of $\beta>1$ the spin parameter $a_{*}$ can take on values larger than 1 without the formation of a naked singularity. 
 
In this paper, we study whether it is possible to test BH solutions in VTG gravity using iron line spectroscopy and place constraints on the parameter $\beta$ and charge $Q$. Previous work in the context of iron line spectroscopy~\cite{Schee:2008fc, Stuchlik:2010zz, Johannsen:2012ng, Bambi:2012at, 2013JCAP...04..005S, Bambi:2013hza, Jiang:2014loa, Jiang:2015dla, Zhou:2016koy, Ni:2016rhz, Cao:2016zbh, Bambi:2016sac, Cao:2017kdq, Tripathi:2018bbu,Wang-Ji:2018ssh,Bambi:2018ggp,super-review} has shown that it can be a very powerful tool for probing the strong gravity regime near BHs. In principle, with high quality data from X-ray telescopes and with the correct astrophysical model to describe the accretion disk, such observations can be used to place stringent constraints on deviations from the Kerr spacetime.

In an effort to determine the constraining power of current X-ray missions to test and constrain VTG gravity, we perform a preliminary study by simulating observations of K$\alpha$ iron lines from accretion disks around VTG BHs and fit the data with Kerr iron lines. If the fits are good, irrespective of whether the physical parameters are well-recovered or not, we argue that the iron line shapes in VTG do not significantly deviate from the iron line shapes in Kerr. Thus, current reflection spectrum observations cannot be used to place constraints on VTG. We find this to be the case in the weak coupling regime of VTG, where the parameter $\beta$ is relatively small, as is the deviation from Kerr. On the other hand, if the fits are bad, we argue that current observations of the reflection spectrum can be used to place constraints on VTG. We determine this is the case in the strong and ultraspinning regimes, where the deviation in the iron line is large enough that deviations in the full reflection spectrum would be detectable with current observations. Of course, the full reflection spectrum is a much more complex model in comparison and a full analysis is required for any definitive conclusion, which we leave for future work.

This paper is organized as follows. In Section~\ref{sec: BH VTG}, we briefly review the black hole solution in VTG gravity found in~\cite{Filippini:2017kov}. In Section~\ref{sec: iron line}, we summarize the iron line method and calculate a set of iron line shapes from putative accretion disks around BHs in VTG. In Section~\ref{sec: sims}, we simulate iron line observations with \textsl{NuSTAR} and determine whether an analysis of the iron line can distinguish between Kerr BHs and the BHs of VTG. Finally, we summarize and conclude in Section~\ref{sec: conc}. Throughout this paper we employ a metric with signature $(-+++)$ and units in which $G_{\text{N}}=c=\hbar=1$, with the exception of Eqs.~(\ref{eq-exc-1}) and (\ref{eq-exc-2}), where we use units in which $4 \pi G_{\text{N}}=c=\hbar=1$.

\section{Black Holes in Vector-Tensor Galileons Modified Gravity}
\label{sec: BH VTG}

The action of VTG is found by starting with the Einstein-Maxwell action
\begin{equation}\label{eq-exc-1}
S_{\text{EM}}=\int d^{4}x\sqrt{-\tilde{g}}\left[\frac{\tilde{R}}{4}-\frac{1}{4}\tilde{F}^{\mu\nu}\tilde{F}_{\mu\nu}\right].
\end{equation}
Note that while the action is the same as that for an Einstein-Maxwell system, the vector field $F_{\mu\nu}$ is associated with an additional dark force rather than standard electromagnetism.

The Einstein-Maxwell action is acted upon by a disformal transformation~\cite{Bekenstein:1992pj, Bettoni:2013diz, Zumalacarregui:2013pma, Kimura:2016rzw} given by
\begin{align}
\tilde{g}_{\mu\nu}(x)&=g_{\mu\nu}(x)-\beta^{2}A_{\mu}(x)A_{\nu}(x),\label{eq: dis trans}
\\
\tilde{A}_{\mu}(x)&=A_{\mu}(x)+\partial_{\mu}\alpha(x),
\end{align}
where $\beta$ is a real constant, $A_{\mu}(x)$ is a vector field, and $\alpha(x)$ is an arbitrary function associated with the gauge freedom of the theory and will not appear going forward.

The disformed action is then, up to total derivatives, given by~\cite{Filippini:2017kov}
\begin{align}\label{eq-exc-2}
S&=\int d^{4}x\sqrt{-g}\frac{1}{4\gamma_{0}}\left[R-\frac{\beta^2}{4}\gamma_{0}^{2}\left(S_{\mu\nu}S^{\mu\nu}-S^{2}\right)\right.
\nonumber \\
&\left.-\frac{4-\beta^{2}}{4}F_{\mu\nu}F^{\mu\nu}+\frac{\beta^{2}-4\beta^{2}}{2}\gamma_{0}^{2}F_{\mu\rho}F_{\nu}^{\rho}A^{\mu}A^{\nu}\right],
\end{align}
with
\begin{align}
\gamma_{0}^{2}&=\frac{1}{1-\beta^{2}A^{\mu}A_{\mu}},
\\
F_{\mu\nu}&=\nabla_{\mu}A_{\nu}-\nabla_{\nu}A_{\mu},
\\
S_{\mu\nu}&=\nabla_{\mu}A_{\nu}+\nabla_{\nu}A_{\mu},
\\
S&=S_{\mu\nu}g^{\mu\nu}.
\end{align}

Finding a BH solution is simply a case of applying the disformal transformation to a suitable solution of the Einstein-Maxwell action. The well-known Kerr-Newman solution that describes a charged, rotating black hole is given by (in Boyer-Lindquist coordinates)
\begin{align}
ds^{2}&=\left(\frac{dr^{2}}{\Delta}+d\theta^{2}\right)\rho^{2}-\left(dt-a\sin^{2}\theta d\phi^{2}\right)^{2}\frac{\Delta}{\rho^{2}}
\nonumber \\
&+\left[\left(r^{2}+a^{2}\right)d\phi-adt\right]^{2}\frac{\sin^{2}\theta}{\rho^{2}},
\end{align}
with
\begin{align}
\Delta&=r^{2}+a^{2}-2Mr+Q^{2},
\\
\rho^{2}&=r^{2}+a^{2}\cos^{2}\theta.
\end{align}
Here, $M$ is the BH mass, $a=|\vec{J}|/M$,  $\vec{J}$ is the BH spin angular momentum, and $Q$ is the BH charge. It is often convenient to introduce the dimensionless spin parameter $a_* = a/M = |\vec{J}|/M^2$.

Applying the disformal transformation of Eq.~(\ref{eq: dis trans}), along with the ansatz $A(r)=Qr/\Delta(r)$, gives the rotating BH solution (in Boyer-Lindquist coordinates)~\cite{Filippini:2017kov}
\begin{align}
ds^{2}&=\left(\frac{\rho^{2}}{\Delta\rho^{2}-\beta^{2}Q^{2}r^{2}}dr^{2}+d\theta^{2}\right)\rho^{2}
\nonumber \\
&-\left(dt-a\sin^{2}\theta d\phi\right)^{2}\frac{\Delta\rho^{2}+\beta^{2}Q^{2}r^{2}}{\rho^{4}}
\nonumber \\
&+\left[\left(r^{2}+a^{2}\right)d\phi-a dt\right]^{2}\frac{\sin^{2}\theta}{\rho^{2}},\label{eq:metric}
\end{align}
The charge $Q$ here is associated with a dark force rather than an electromagnetic charge as in the Kerr-Newman solution. Note that the dimensionless parameter $\beta$ controls the deviation away from the Kerr-Newman solution. For $\beta=0$, Eq.~(\ref{eq:metric}) reduces to the Kerr-Newman solution and for $Q=0$ it reduces to the Kerr solution.

The event horizon can be found by solving the equation $\partial^{\mu}r\partial_{\nu}r=g^{rr}=0$. This can be reduced to finding the real positive solutions of~\cite{Vetsov:2018mld}
\begin{align}
&r^{4}-2Mr^{3}+\left[a^{2}\left(1+\cos^{2}\theta\right)-\sigma M^{2}\right]r^{2}
\nonumber \\
&-2a^{2}M\cos^{2}\theta r+a^{2}\cos^{2}\theta\left(a^{2}+\frac{\sigma Q^{2}}{1-\beta^{2}}\right)=0,
\end{align}
where
\begin{equation}
\sigma=\frac{Q^{2}}{M^{2}}\left(\beta^{2}-1\right).
\end{equation}
This is a fourth order algebraic equation in the radial coordinate $r$ and can have four, two, or no real roots depending on the sign of the determinant. The maximal real root corresponds to the position of the event horizon and it has some interesting properties. 

In contrast to the horizon in the Kerr-Newman solution given by
\begin{equation}
r^{\text{KN}}_{\text{H}}=M+\sqrt{M^{2}-a^{2}-Q^{2}},
\end{equation}
the horizon in the disformal solution is not spherically symmetric, as it depends on the polar angle $\theta$. 

Additionally, while the Kerr-Newman solution requires that $a \leq \sqrt{M^{2}-Q^{2}}\leq M$ for the existence of a horizon, this is not necessarily the case in the disformal solution. This can be easily seen from the horizon radius in the equatorial plane of the disformal solution given by
\begin{equation}
r_{\text{H}}(\theta=\pi/2)=M+\sqrt{M^{2}-a^{2}-Q^{2}(1-\beta^{2})}.
\end{equation}
For values of $\beta>1$, the spin $a$ can take on values larger than $M$ and a horizon will still exist. The maximum value of $a$ is given by
\begin{equation}
a^{\text{max}}=\frac{M}{2}\sqrt{2-\sigma+2\sqrt{1-\sigma}}.
\end{equation}

Another interesting property of the event horizon in the disformal solution is that for values of $\beta>1$ the horizon exists even in the massless limit $M\rightarrow0$. In this case the horizon radius is given by
\begin{align}
&r_{\text{H}}^{2}=\frac{1}{2}\left[Q^{2}(\beta^{2}-1)-a^{2}(1+\cos^{2}\theta)\right.
\nonumber \\
&\left.+\sqrt{\left[Q^{2}(\beta^{2}-1)-a^{2}(1+\cos^{2}\theta)\right]^{2}-4a^{2}(a^{2}+Q^{2})\cos^{2}\theta}\right].
\end{align}

Further discussion of the horizon can be found in Refs.~\cite{Filippini:2017kov} and~\cite{Vetsov:2018mld}.

\section{Iron Line Spectroscopy}
\label{sec: iron line}

We consider a system containing a central BH accreting from a geometrically thin and optically thick disk~\cite{Bambi:2017iyh, Bambi:2017khi}. The disk emits locally as a blackbody and when integrated radially the spectrum is a multi-temperature blackbody. The temperature of the gas making up the disk is a function of the properties of the BH,~e.g.~the BH mass and spin, the accretion rate, and the radial distance from the center of the BH. For an accretion rate of $\sim 10\%$ of the Eddington limit, the temperature of the innermost part of the disk is in the soft X-ray band,~i.e.~$\sim1$ keV, for stellar-mass BHs and in the optical/UV bands,~i.e.~$1-10$ eV, for supermassive BHs.

In addition to the accretion disk, we include within the model a hotter ($\sim 100$ keV), usually optically thin, cloud of gas near the BH termed a \textit{corona}. The exact geometry of the corona in BH-accretion disk systems is not known at the moment, but several models have been proposed. For example, the corona could be the base of a jet, the atmosphere just above/below the accretion disk, or some accretion flow near the BH. In general, though, the corona plays an important part in the electromagnetic radiation of the system. Thermal photons from the disk can inverse Compton scatter off free electrons in the corona, generating a power-law spectrum with an energy cut-off (that is dependent on the corona temperature). The photons of this power-law component can, in turn, illuminate the disk, producing a reflection spectrum. The most prominent feature of the reflection spectrum is usually the iron K$\alpha$ line, and it is also the feature providing the most information on the spacetime metric describing the BH, particularly the strong gravity region near the BH. Tests of the strong gravity region using real X-ray data require fitting the entire reflection spectrum, not just the iron line~\cite{Bambi:2016sac, Cao:2017kdq, Tripathi:2018bbu}, but in this work, as a preliminary and exploratory study, we will only consider the iron K$\alpha$ line.

The iron K$\alpha$ line is a very narrow line feature when emitted, but becomes broadened and skewed when observed at large distances due to the combined emission from different regions of the accretion disk and the effects of gravitational redshift and Doppler boosting. The line is emitted at $6.4$ keV in the case of neutral or weakly-ionized iron and can shift up to $6.97$ keV in the case of hydrogen-like iron ions.

The shape of the iron line as observed far from the source is dependent on the spacetime geometry of the BH, the inclination angle between the angular momentum of the disk and the observer's line of sight, the geometry of the emission region, and the emissivity profile. The BH metric in VTG studied in this paper depends on the spin parameter $a$, the charge $Q$, and the dimensionless parameter $\beta$ (the BH mass $M$ does not impact the shape of the iron line). The inclination angle $\iota$ ranges from $0\degree$ (face-on disk) to $90\degree$ (edge-on disk). We set the inner edge of the accretion disk equal to the innermost stable circular orbit (ISCO), which depends on $a$, $Q$, and $\beta$, and the outer edge is set at a sufficiently large radius such that slightly altering its value does not significantly impact the iron line. The intensity profile of the disk is modeled with a simple power-law of the form $1/r^{q}$, where $q$ is the emissivity index.

To calculate the iron line as measured by a distant observer we begin from the photon number count far from the source
\begin{align}
N(E_{\text{obs}})&=\frac{1}{E_{\text{obs}}}\int I_{\text{obs}}(E_{\text{obs}})\frac{dXdY}{D^{2}}
\nonumber \\
&=\frac{1}{E_{\text{obs}}}\int g^{3}I_{\text{e}}(E_{\text{e}})\frac{dXdY}{D^{2}},\label{eq: photon}
\end{align}
where $E_{\text{obs}}$ and $I_{\text{obs}}$ are the photon energy and specific intensity of the radiation at the observer, respectively, $E_{\text{e}}$ and $I_{\text{e}}$ are the same quantities at the emission point, $g=E_{\text{obs}}/E_{\text{e}}$ is the redshift factor, and $I_{\text{obs}}=g^{3}I_{\text{e}}$ follows from Liouville's theorem~\cite{1973grav.book}. $X$ and $Y$ are the Cartesian coordinates on the image plane of the observer, and $D$ is the distance between the BH and the observer. Assuming monochromatic emission with a power-law profile the emitted intensity is given by
\begin{equation}
I_{\text{e}}(E_{\text{e}})\propto \frac{\delta(E_{\text{e}}-E_{*})}{r^{q}},
\end{equation}
where $E_{*}=6.4$ keV.

We compute Eq.~(\ref{eq: photon}) by using the ray-tracing code described in~\cite{Bambi:2012tg}. This code solves the equations of motion for photons on a given spacetime background. The photons are initialized with a three-momentum perpendicular to the image plane of the observer and are then evolved backwards in time to determine their emission point in the accretion disk. Particles of gas within the accretion disk follow nearly geodesic circular orbits in the equatorial plane of the BH. The photon four-velocity can be written as $u_{\text{e}}^{\mu}=u_{\text{e}}^{t}(1,0,0,\Omega)$, where $\Omega=u_{\text{e}}^{\phi}/u_{\text{e}}^{t}$ is the Keplerian angular velocity. In terms of the metric elements the angular velocity is given by
\begin{equation}
\Omega_{\pm}=\frac{(-\partial_{r}g_{t\phi})\pm\sqrt{(\partial_{r}g_{t\phi})^{2}-(\partial_{r}g_{tt})(\partial_{r}g_{\phi\phi})}}{\partial_{r}g_{\phi\phi}},
\end{equation}
where $+(-)$ refers to corrotating (counterrotating) orbits,~i.e.~orbits with angular momentum parallel (anti-parallel) to the black hole spin angular momentum. From the normalization condition $g_{\mu\nu}u_{\text{e}}^{\mu}u_{\text{e}}^{\nu}=-1$ we have
\begin{equation}
u_{\text{e}}^{t}=\frac{1}{\sqrt{-g_{tt}-2g_{t\phi}\Omega-g_{\phi\phi}\Omega^{2}}}.
\end{equation}
The redshift factor is
\begin{equation}
g=\frac{-u_{\text{obs}}^{\mu}k_{\mu}}{-u_{\text{e}}^{\nu}k_{\nu}},
\end{equation}
where $k^{\mu}$ is the photon four-momentum and $u_{\text{obs}}^{\mu}=(1,0,0,0)$ is the four-velocity of the distant observer. The redshift factor can be rewritten in terms of the metric
\begin{equation}
g=\frac{\sqrt{-g_{tt}-2g_{t\phi}\Omega-g_{\phi\phi}\Omega^{2}}}{1+\lambda\Omega},
\end{equation}
where $\lambda=k_{\phi}/k_{t}$ is a constant of motion along the photon's path and can be determined from the photon's initial conditions. For more details on the calculations of the spectrum of thin disks around black holes see~\cite{Johannsen:2012ng, Bambi:2012at, Bambi:2016sac, Bambi:2017khi}.

\begin{figure*}[hpt]
\includegraphics[width=\columnwidth,clip]{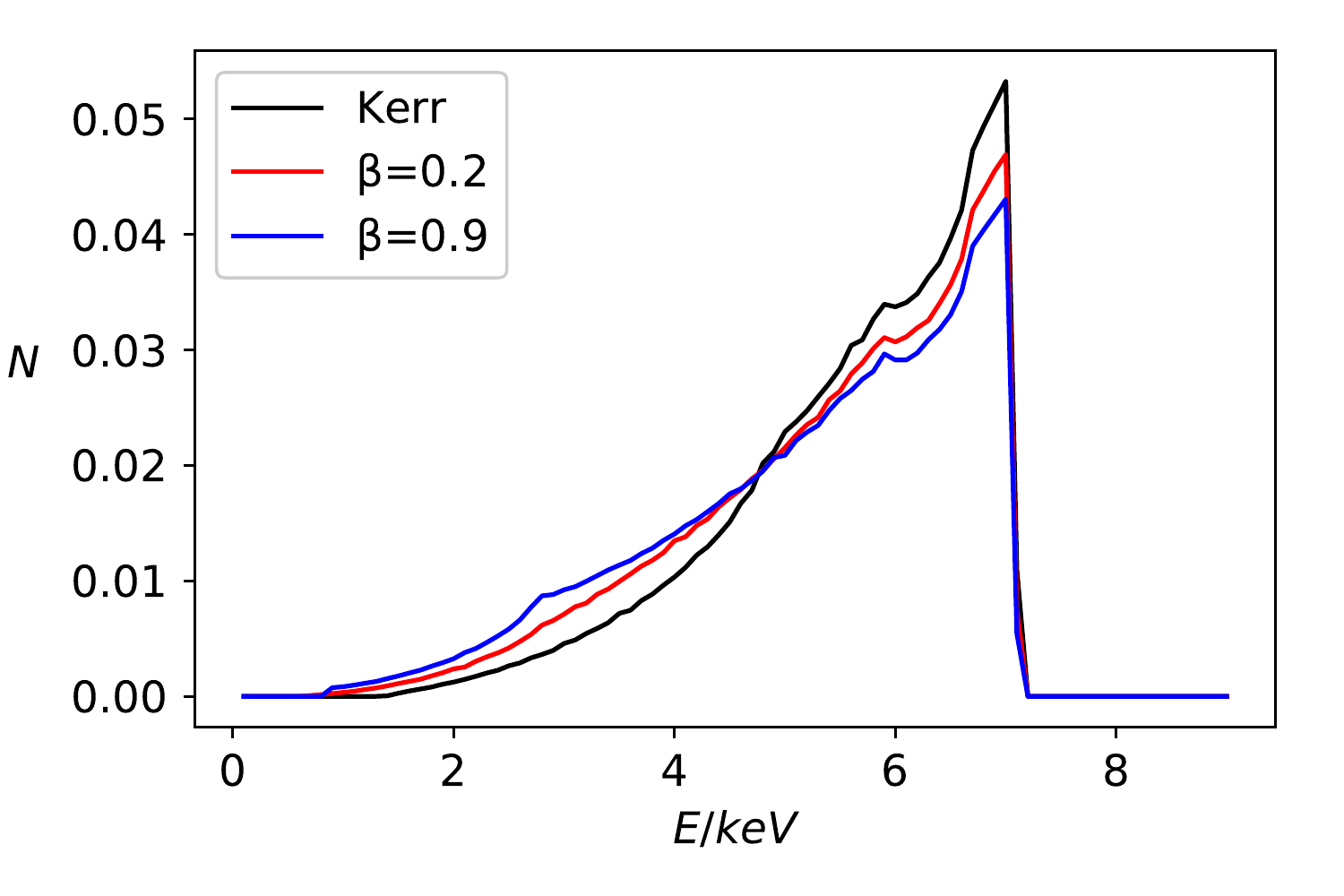}
\caption{\label{fig: weak} Iron line shapes in VTG in the weak coupling regime,~i.e.~$0<\beta<1$, for spin parameter $a_{*}=0.9$, inclination angle $\iota=45\degree$, charge $Q=0.4$, and emissivity index $q=3$.}
\end{figure*}

\begin{figure*}[hpt]
\includegraphics[width=\columnwidth,clip]{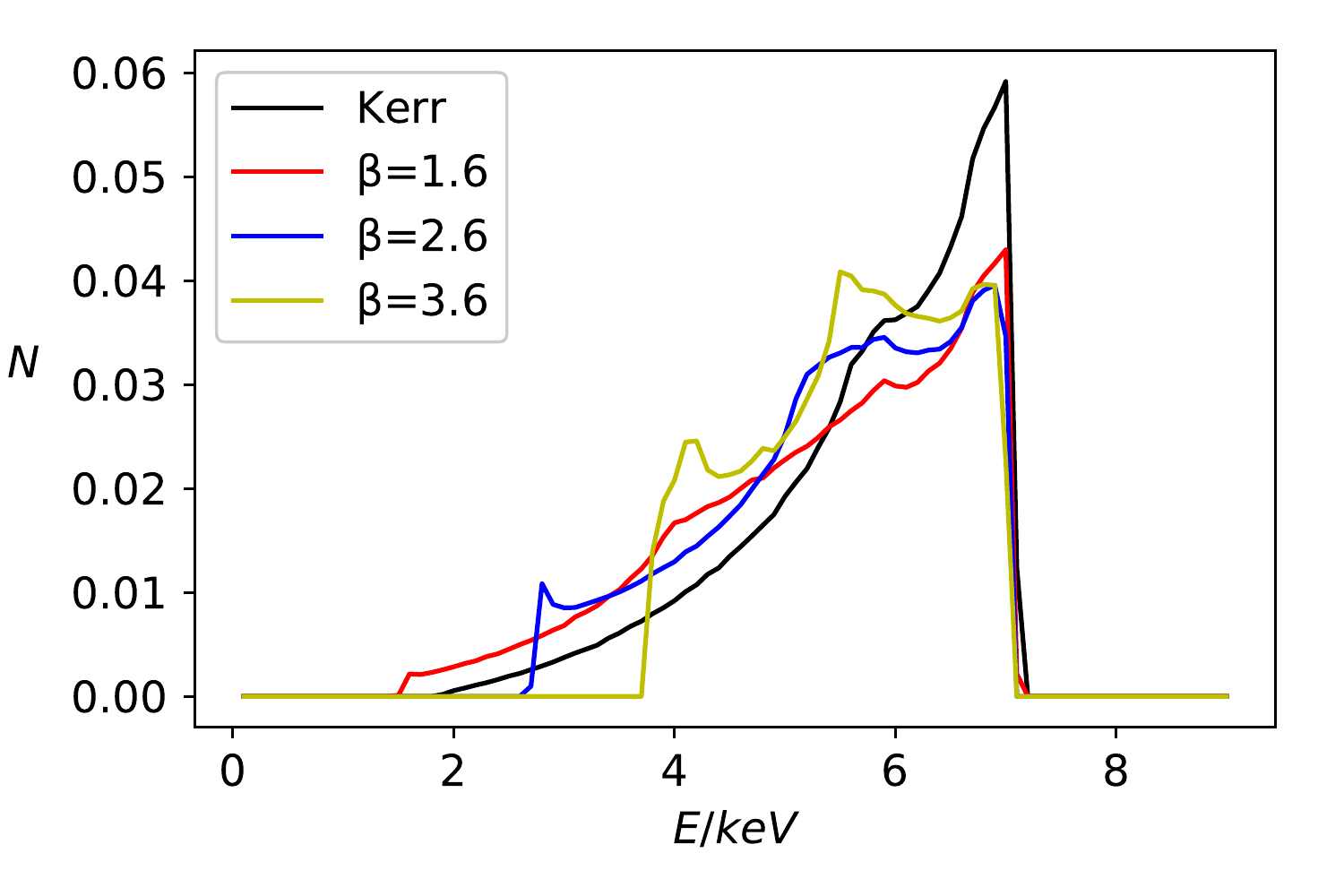}
\includegraphics[width=\columnwidth,clip]{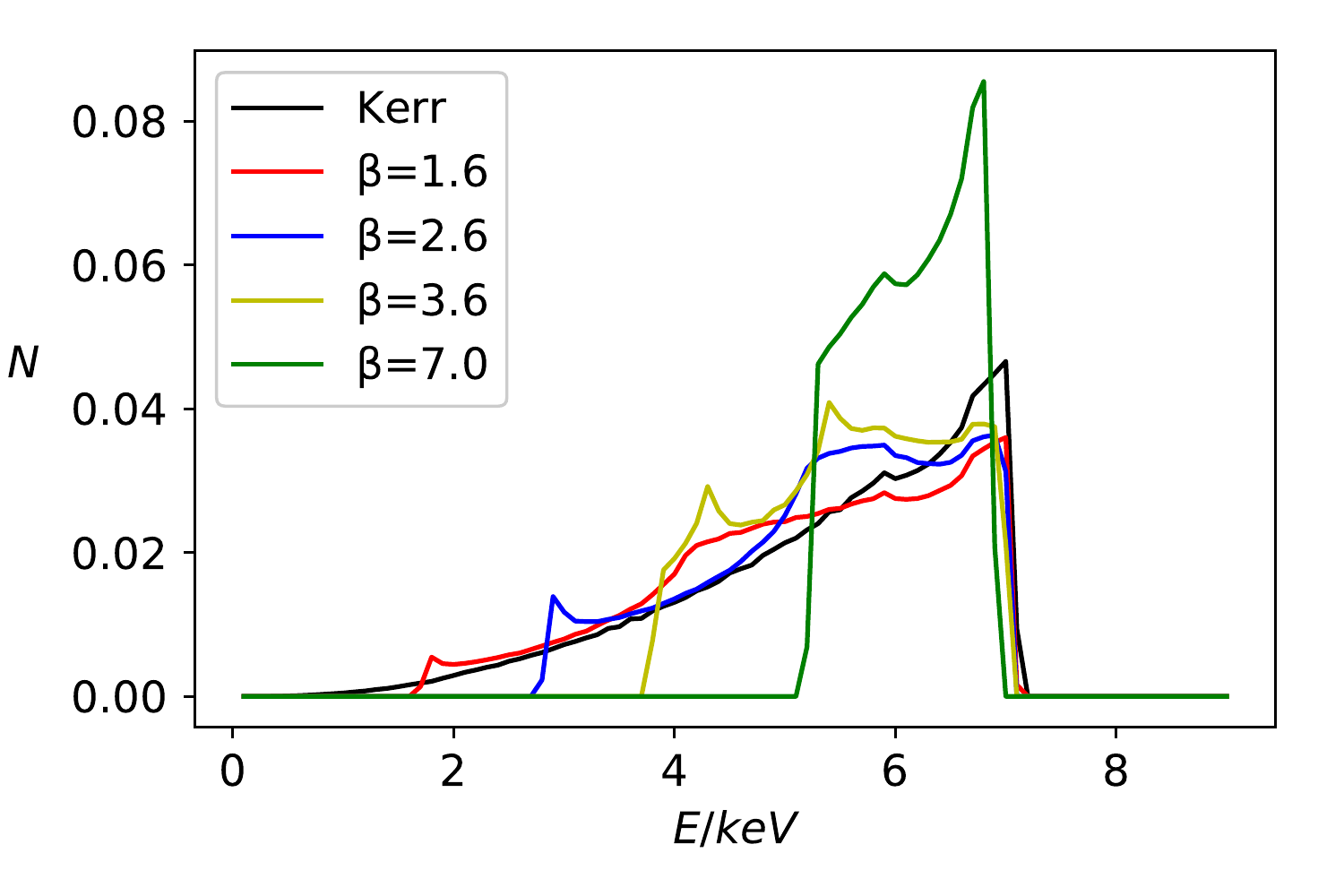}
\caption{\label{fig: strong} Iron line shapes in VTG in the strong coupling regime,~i.e.~$\beta>1$, for spin parameters $a_{*}=0.8$ (left) and $0.995$ (right), inclination angle $\iota=45\degree$, charge $Q=0.4$, and emissivity index $q=3$.}
\end{figure*}

\begin{figure*}[hpt]
\includegraphics[width=\columnwidth,clip]{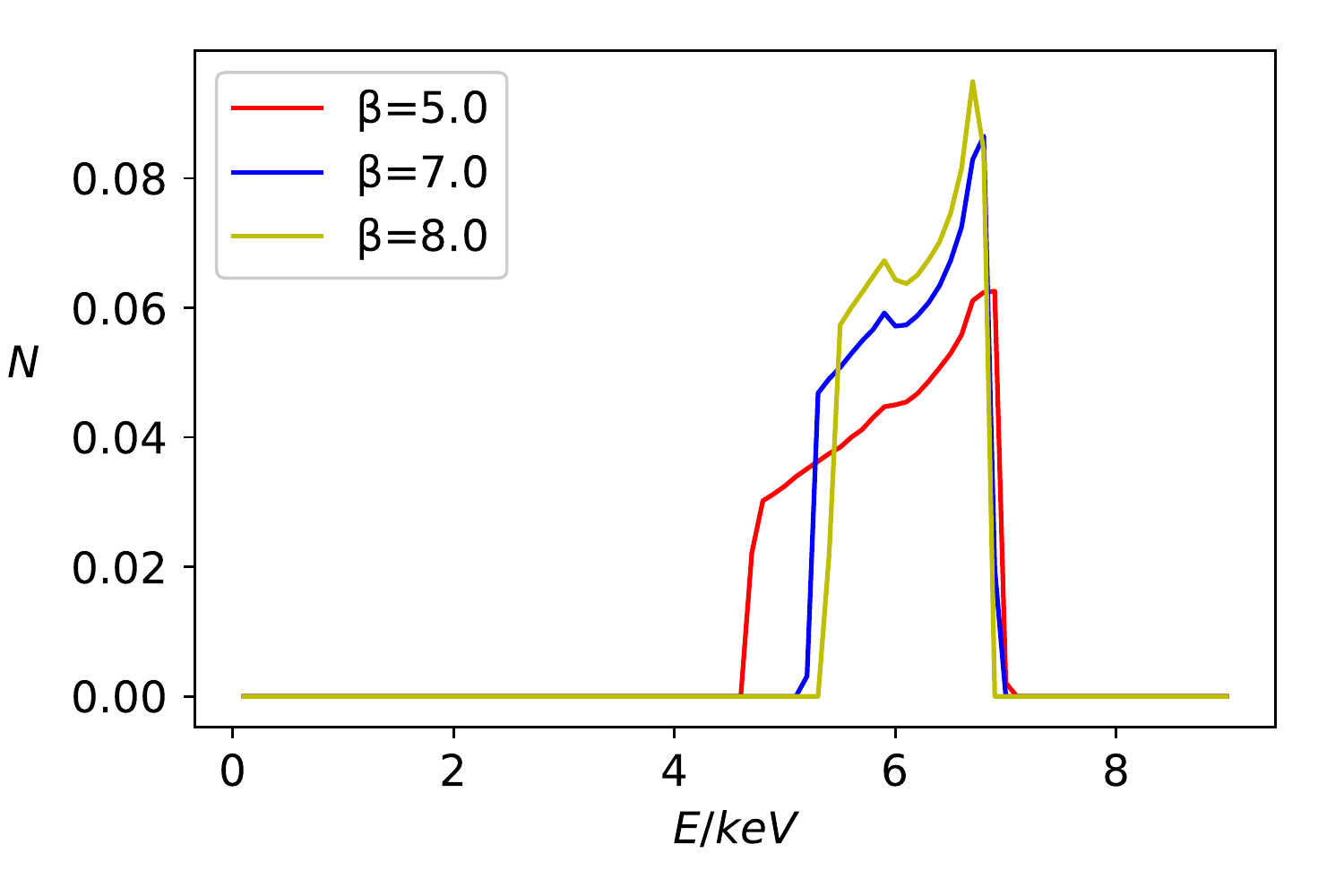}
\includegraphics[width=\columnwidth,clip]{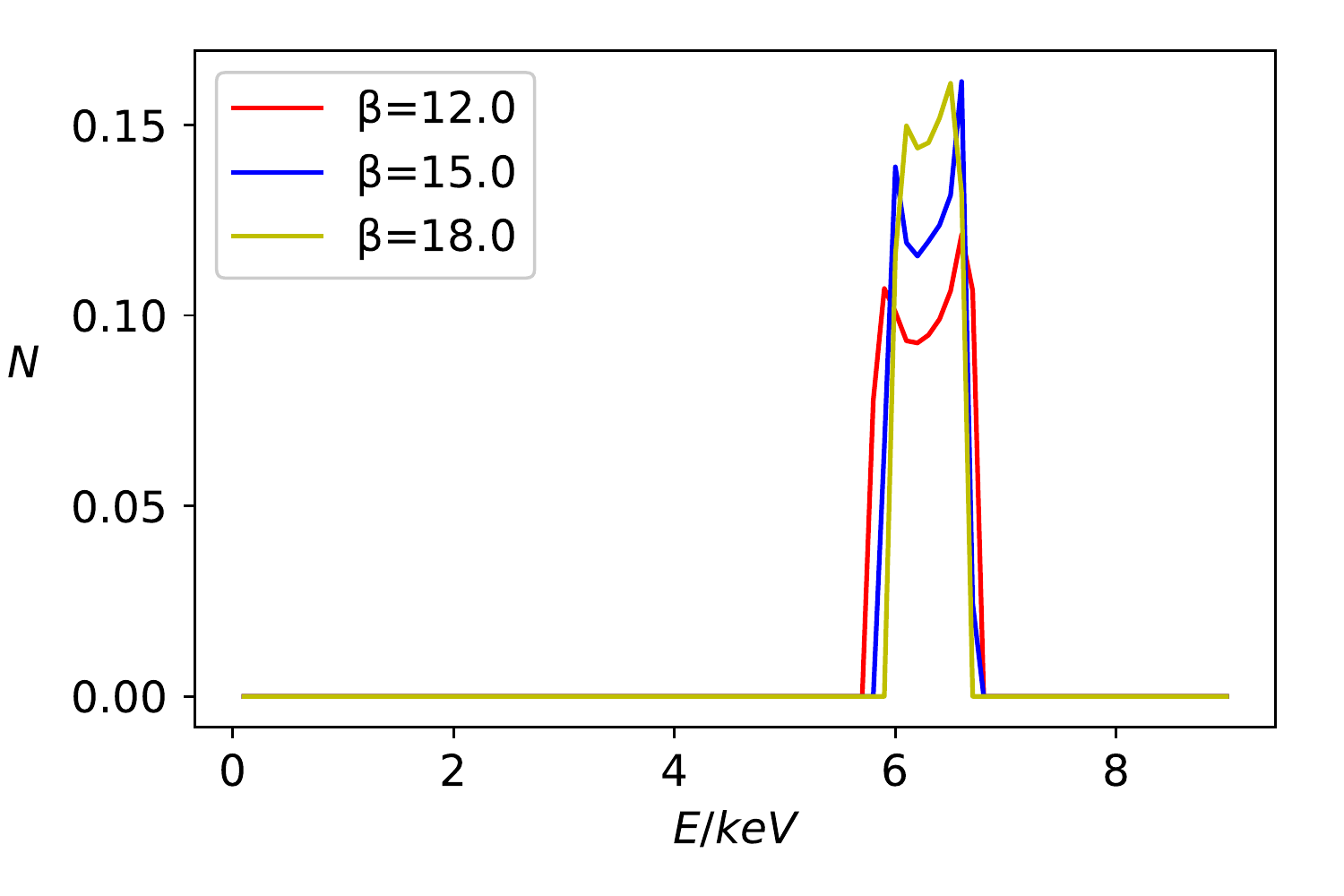}
\\
\includegraphics[width=\columnwidth,clip]{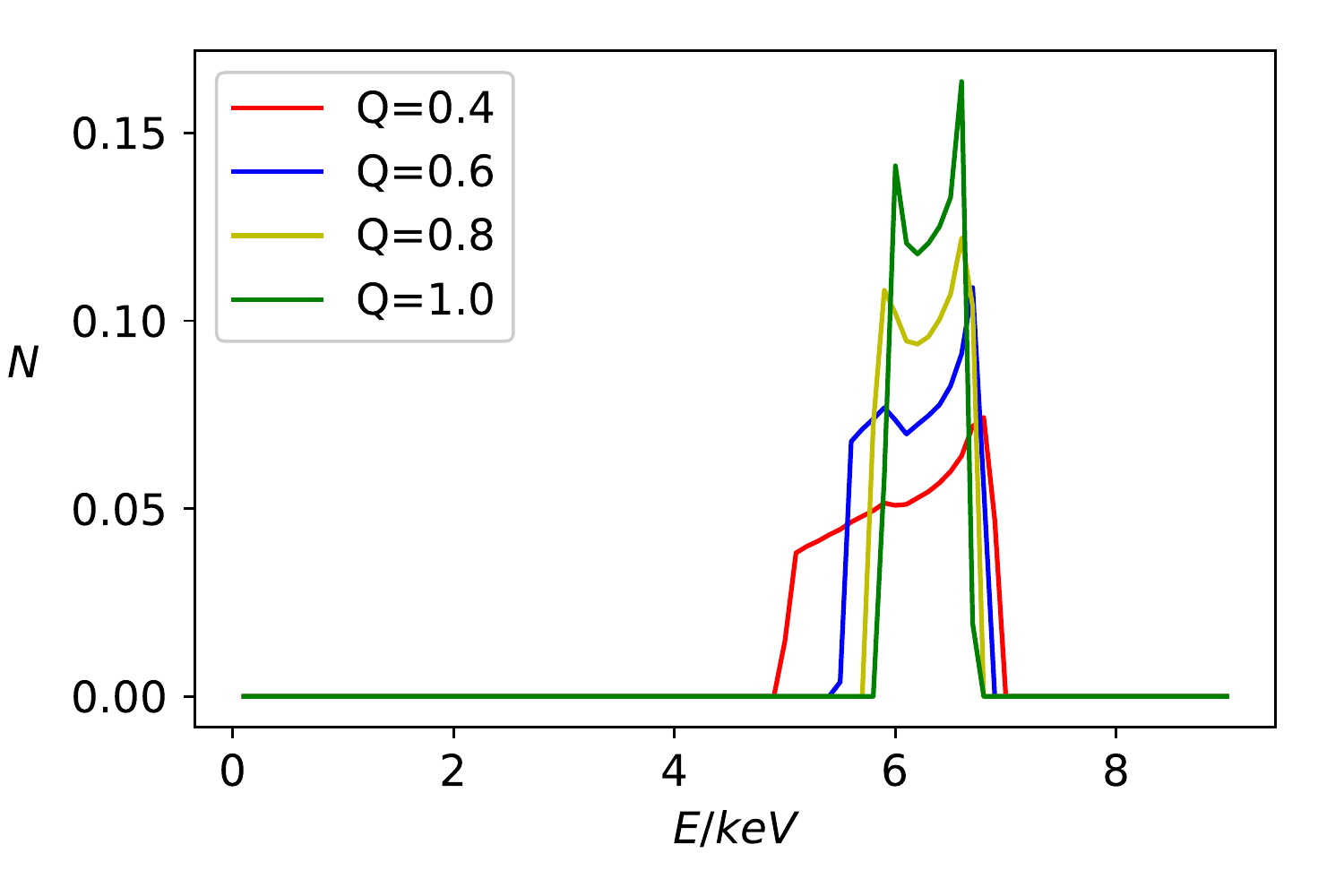}
\caption{\label{fig: ultra} Iron line shapes in VTG in the ultraspinning regime,~i.e.~$\beta>1$ and $a_*>1$, for spin parameters $a_{*}=2.0$ (top left and bottom) and $4.0$ (top right), inclination angle $\iota=45\degree$, and emissivity index $q=3$. The charge $Q=0.4$ in the top panels and the coupling parameter $\beta=6.0$ in the bottom panel.}
\end{figure*}

Iron K$\alpha$ line shapes of the reflection spectrum of BHs in VTG gravity are shown in Figs.~\ref{fig: weak}--\ref{fig: ultra}. Figure~\ref{fig: weak} shows the iron line in the weak coupling regime,~i.e.~$0<\beta<1$, for spin parameter $a_{*}=0.9$ and charge $Q=0.4$. Figure~\ref{fig: strong} shows the iron line in the strong coupling regime,~i.e.~$\beta>1$, for spin parameters $a_{*}=0.8$ and $0.998$ and charge $Q=0.4$. Figure~\ref{fig: ultra} shows the iron line in the ultraspinning regime,~i.e.~$\beta>1$ and $a_{*}>1$, for spin parameters $a_{*}=2.0$ and $4.0$ and an assortment of values for coupling parameter $\beta$ and charge $Q$. For all figures the inclination angle $\iota=45\degree$ and the emissivity index $q=3$. The iron line in Kerr is shown in black for comparison in the figures for the weak and strong coupling regimes. The iron lines in the weak coupling regime do not show significant deviation from the Kerr line, while there is significant deviation in the strong coupling regime, although the general shape of the lines is still similar to that seen in Kerr except for the largest values of $\beta$. In the ultraspinning regime, the iron line shapes depart significantly from those in Kerr, particularly for larger values of $\beta$ and $Q$.

%

\section{Simulations}
\label{sec: sims}

The primary goal of this work is to determine if observations from current X-ray missions can be used to test VTG gravity and constrain the charge $Q$ and the parameter $\beta$. As stated previously, we will not be constructing the full reflection spectrum model to analyze real data of specific sources. Instead, as a preliminary and exploratory study, we follow the strategy employed in previous studies~\cite{Zhou:2016koy, Ni:2016rhz, Cao:2016zbh, Zhang:2018xzj}. We simulate observations of a VTG BH iron line using the response characteristics of a current X-ray mission and fit the simulated data with a Kerr iron line. If the fit is found to be good, regardless of whether the fit parameters agree with the simulated parameters, we argue that the iron line in VTG is not sufficiently different from that of Kerr to use observations of the reflection spectrum with current telescopes to test VTG. On the other hand, if the fit is bad, we argue that VTG can be tested using the iron line method with current X-ray telescopes, and it would be useful to construct the full reflection model to fit real data as done in~\cite{Bambi:2016sac, Cao:2017kdq, Tripathi:2018bbu}.

We perform our analysis on the 19 VTG cases depicted in Figs.~\ref{fig: weak}--\ref{fig: ultra} and discussed in Sec.~\ref{sec: iron line}. Tables~\ref{tab: weak}--\ref{tab: ultra-2} summarize the parameters used in each of the analyzed simulations. We choose \textsl{NuSTAR} as the representative current X-ray mission and use its response files downloaded from the \textsl{NuSTAR} website\footnote{http://www.nustar.caltech.edu}. We use the software \textsc{Xspec} to simulate and fit the data\footnote{http://heasarc.gsfc.nasa.gov/docs/xanadu/xspec/index.html}. We consider typical parameters of a bright black hole binary. The simulated spectra are generated assuming a simple power-law with photon index $\Gamma=1.6$ (modeling the power-law spectrum of the corona) and a single iron line (modeling the reflection spectrum of the accretion disk). The luminosity of the source is set at $10^{-9}$ erg/s/cm$^{2}$ in the band $3-10$ keV. The equivalent width of the iron line is around 200 eV. We consider both instruments onboard \textsl{NuSTAR},~i.e.~FPMA and FPMB, and observations of 200 ks in length.

\begin{table*}[hpt]
\begin{center}
\begin{tabular}{l c c}
\hline \hline
Sim & 1 & 2 \\
Input \\
$a_{*}$ & 0.9 & 0.9 \\
$\beta$ & 0.2 & 0.9 \\ 
Q & 0.4 & 0.4 \\ \hline
Best-fit \\
$a_{*}$ & $>0.734$ & $>0.803$ \\
$\iota$ [deg] & $45.6^{+1.4}_{-1.3}$ & $44.4\pm{1.3}$ \\
q & $3.67^{+0.41}_{-0.70}$ & $3.30^{+0.32}_{-0.30}$ \\
$\Gamma$ & 1.60 & 1.61 \\ \hline
$\chi_{\text{min}}^{2}$ & 1.07 & 0.98 \\ \hline \hline
\end{tabular}
\caption{\label{tab: weak} Summary of the best-fit values of simulations in the weak coupling regime,~i.e.~$0<\beta<1$. In both simulations the inclination angle $\iota=45\degree$, the emissivity index $q=3$, and the photon index $\Gamma=1.6$. The reported uncertainty is at the $90\%$ confidence level for one relevant parameter. The uncertainty for $\Gamma$ is always less than 0.01.}
\end{center}
\end{table*}
\begin{figure*}[hpt]
\includegraphics[width=\columnwidth,clip]{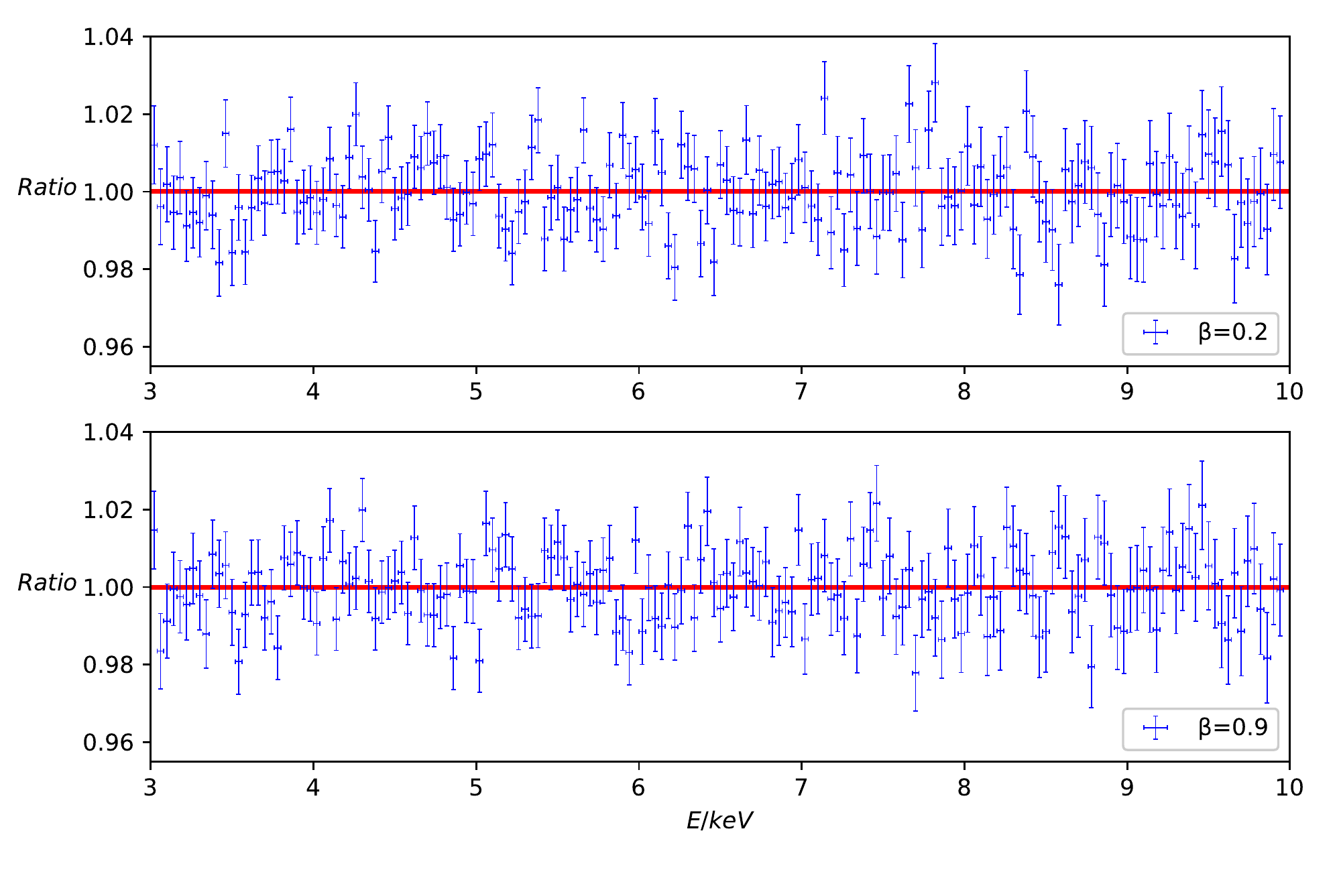}
\caption{\label{fig: rat-weak} Ratio between data and the best-fit model for simulations 1 and 2 in the weak coupling regime,~i.e.~$0<\beta<1$, of VTG. See text and Table~\ref{tab: weak} for more details.}
\end{figure*}
\begin{table*}[hpt]
\begin{center}
\begin{tabular}{l c c c c c c c}
\hline \hline
Sim & 3 & 4 & 5 & 6 & 7 & 8 & 9 \\
Input \\
$a_{*}$ & 0.8 & 0.8 & 0.8 & 0.995 & 0.995 & 0.995 & 0.995 \\
$\beta$ & 1.6 & 2.6 & 3.6 & 1.6 & 2.6 & 3.6 & 7.0 \\
Q & 0.4 & 0.4 & 0.4 & 0.4 & 0.4 & 0.4 & 0.4 \\ \hline
Best-fit \\
$a_{*}$ & $>0.930$ & $0.854^{+0.035}_{-0.044}$ & $0.941^{+0.014}_{-0.016}$ & $>0.919$ & $0.838^{+0.036}_{-0.039}$ & $0.942^{+0.015}_{-0.018}$ & $0.668^{+0.058}_{-0.067}$ \\
$\iota$ [deg] & $45.9^{+1.5}_{-1.3}$ & $44.6^{+2.0}_{-1.8}$ & $43.5\pm{1.0}$ & $46.5^{+1.5}_{-1.4}$ & $44.5^{+2.0}_{-1.7}$ & $43.3\pm{1.0}$ & $38.4\pm{0.6}$ \\
q & $3.53^{+0.25}_{-0.21}$ & $3.75^{+0.41}_{-0.34}$ & $3.64^{+0.14}_{-0.13}$ & $3.74^{+0.27}_{-0.25}$ & $3.96^{+0.42}_{-0.33}$ & $3.73^{+0.15}_{-0.13}$ & $2.78^{+0.12}_{-0.11}$ \\
$\Gamma$ & 1.60 & 1.60 & 1.59 & 1.60 & 1.60 & 1.59 & 1.59 \\ \hline
$\chi_{\text{min}}^{2}$ & 1.07 & 1.12 & 0.94 & 0.88 & 1.16 & 0.90 & 1.35 \\ \hline \hline
\end{tabular}
\caption{\label{tab: strong} Summary of the best-fit values of simulations in the strong coupling regime,~i.e.~$\beta>1$. In all simulations the inclination angle $\iota=45\degree$, the emissivity index $q=3$, and the photon index $\Gamma=1.6$. The reported uncertainty is at the $90\%$ confidence level for one relevant parameter. The uncertainty for $\Gamma$ is always less than 0.01.}
\end{center}
\end{table*}
\begin{figure*}[hpt]
\includegraphics[width=\columnwidth,clip]{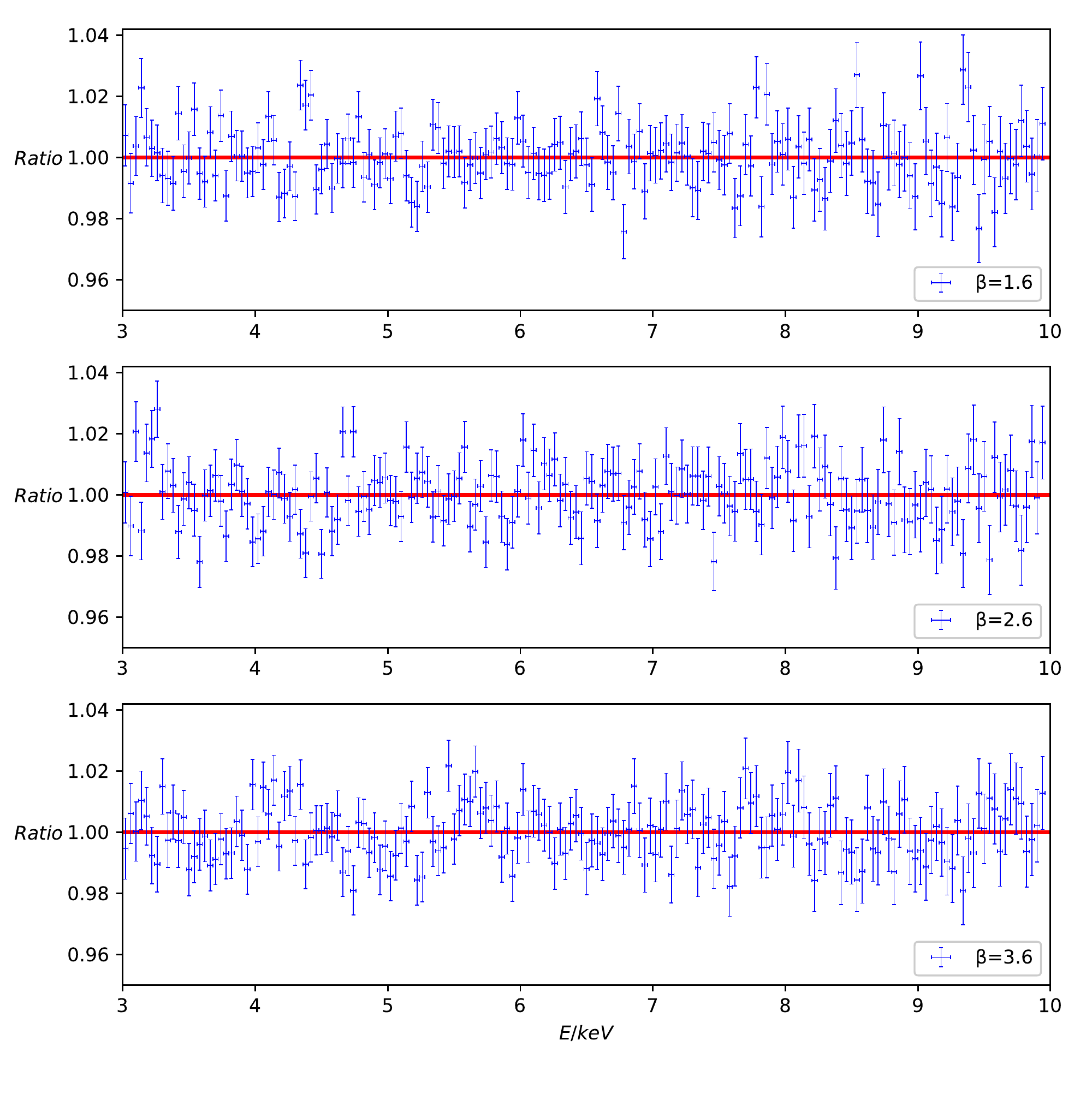}
\includegraphics[width=\columnwidth,clip]{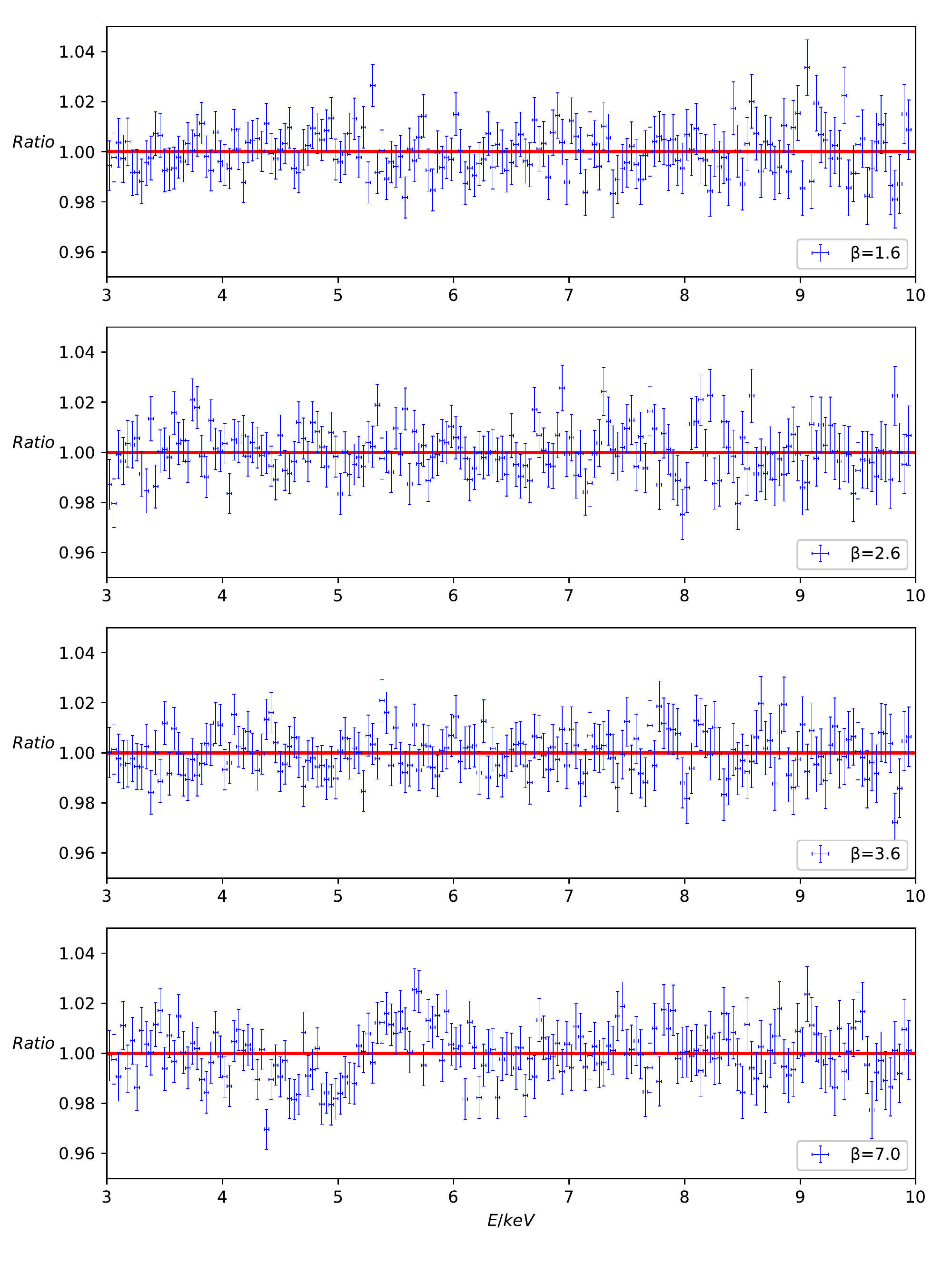}
\caption{\label{fig: rat-strong} Ratio between data and the best-fit model for simulations 3--9 in the strong coupling regime,~i.e.~$\beta>1$, of VTG. The left column is for simulations 3--5 with spin parameter $a_{*}=0.8$ and the right column is for simulations 6--9 with spin parameter $a_{*}=0.995$. See text and Table~\ref{tab: strong} for more details.}
\end{figure*}
\begin{table*}[hpt]
\begin{center}
\begin{tabular}{l c c c c c c c c c c}
\hline \hline
Sim & 10 & 11 & 12 & 13 & 14 & 15 \\
Input \\
$a_{*}$ & 2.0 & 2.0 & 2.0 & 4.0 & 4.0 & 4.0 \\
$\beta$ & 5.0 & 7.0 & 8.0 & 12.0 & 15.0 & 18.0 \\ 
Q & 0.4 & 0.4 & 0.4 & 0.4 & 0.4 & 0.4 \\ \hline
Best-fit \\
$a_{*}$ & $0.798^{+0.041}_{-0.061}$ & $0.668^{+0.057}_{-0.070}$ & $0.599^{+0.081}_{-0.095}$ & $>-0.937$ & $<0.919$ & $<0.991$ \\
$\iota$ [deg] & $40.4\pm{0.8}$ & $38.9^{+0.6}_{-0.7}$ & $37.8^{+0.6}_{-0.7}$ & $36.0\pm1.1$ & $35.2^{+1.6}_{-1.1}$ & $32.4^{+2.1}_{-1.5}$ \\
q & $3.02^{+0.17}_{-0.13}$ & $2.69^{+0.12}_{-0.11}$ & $2.43^{+0.10}_{-0.09}$ & $1.83\pm0.13$ & $1.42^{+0.12}_{-0.10}$ & $0.97^{+0.17}_{-0.30}$ \\
$\Gamma$ & 1.60 & 1.59 & 1.60 & 1.60 & 1.60 & 1.60 \\ \hline
$\chi_{\text{min}}^{2}$ & 1.09 & 1.45 & 1.51 & 2.04 & 1.32 & 1.33 \\ \hline \hline
\end{tabular}
\caption{\label{tab: ultra} Summary of the best-fit values of simulations in the ultraspinning regime,~i.e.~$\beta>1$ and $a_{*}>1$. In all simulations the inclination angle $\iota=45\degree$, the emissivity index $q=3$, and the photon index $\Gamma=1.6$. The reported uncertainty is at the $90\%$ confidence level for one relevant parameter. The uncertainty for $\Gamma$ is always less than 0.01. Note that for simulation 13 the fit was too poor to use the err routine in \textsc{Xspec} to calculate the uncertainties so we report the uncertainties from the steppar routine.} 
\end{center}
\end{table*}
\begin{table*}[hpt]
\begin{center}
\begin{tabular}{l c c c c c c c c c c}
\hline \hline
Sim & 16 & 17 & 18 & 19 \\
Input \\
$a_{*}$ & 2.0 & 2.0 & 2.0 & 2.0 \\
$\beta$ & 6.0 & 6.0 & 6.0 & 6.0 \\ 
Q & 0.4 & 0.6 & 0.8 & 1.0 \\ \hline
Best-fit \\
$a_{*}$ & $0.746^{+0.042}_{-0.051}$ & $0.579^{+0.088}_{-0.108}$ & $0.340^{+0.165}_{-0.211}$ & $<0.936$ \\
$\iota$ [deg] & $39.7^{+0.8}_{-0.7}$ & $37.6^{+0.7}_{-1.0}$ & $33.8^{+1.0}_{-0.8}$ & $33.6^{+0.9}_{-1.3}$ \\
q & $2.96^{+0.16}_{-0.13}$ & $2.31^{+0.11}_{-0.08}$ & $1.97^{+0.17}_{-0.21}$ & $1.43^{+0.14}_{-0.10}$ \\
$\Gamma$ & 1.59 & 1.60 & 1.60 & 1.60 \\ \hline
$\chi_{\text{min}}^{2}$ & 1.24 & 1.89 & 1.76 & 1.10 \\ \hline \hline
\end{tabular}
\caption{\label{tab: ultra-2} Continuation of Table~\ref{tab: ultra}.} 
\end{center}
\end{table*}
\begin{figure*}[hpt]
\includegraphics[width=\columnwidth,clip]{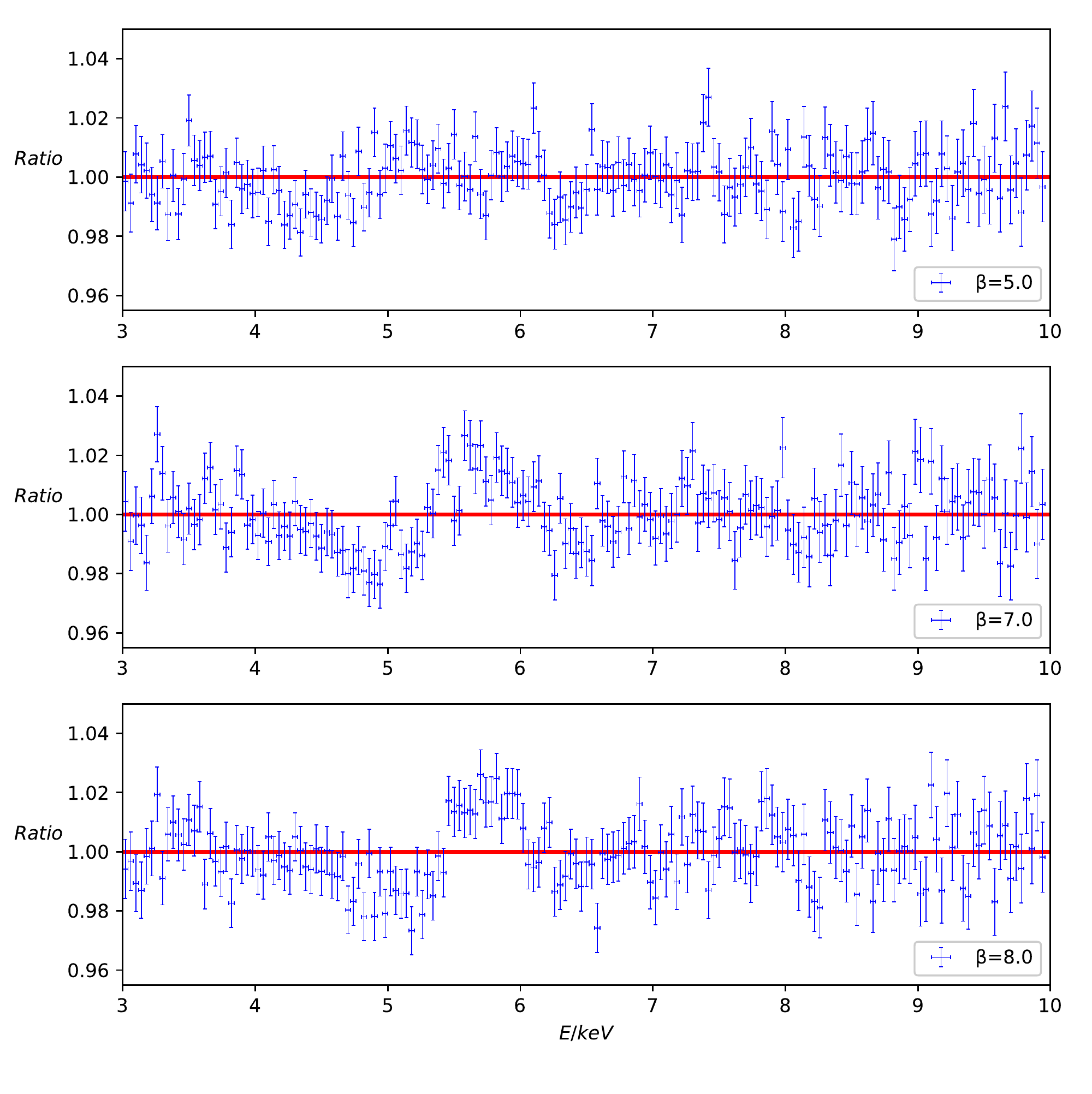}
\includegraphics[width=\columnwidth,clip]{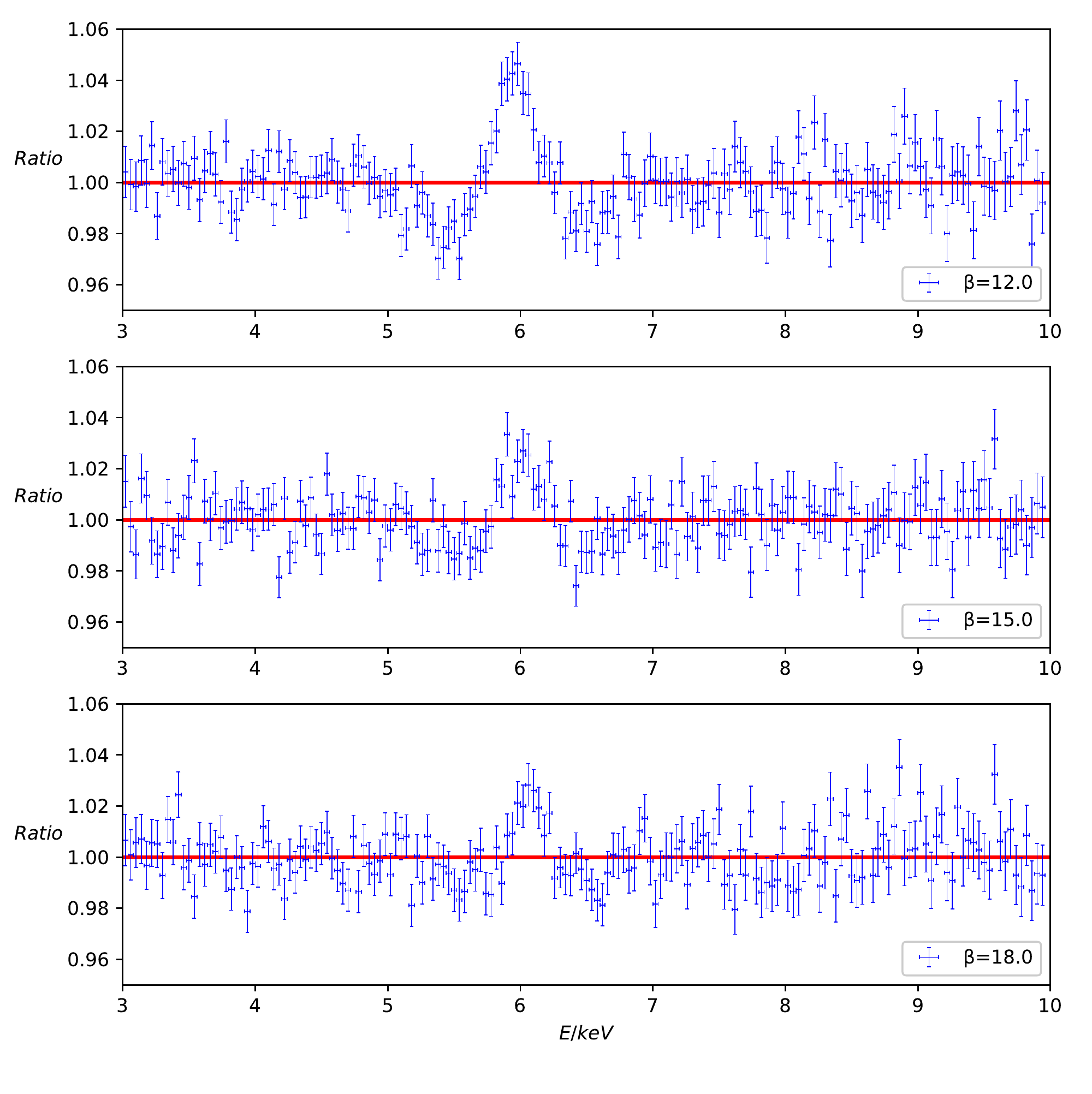}
\includegraphics[width=\columnwidth,clip]{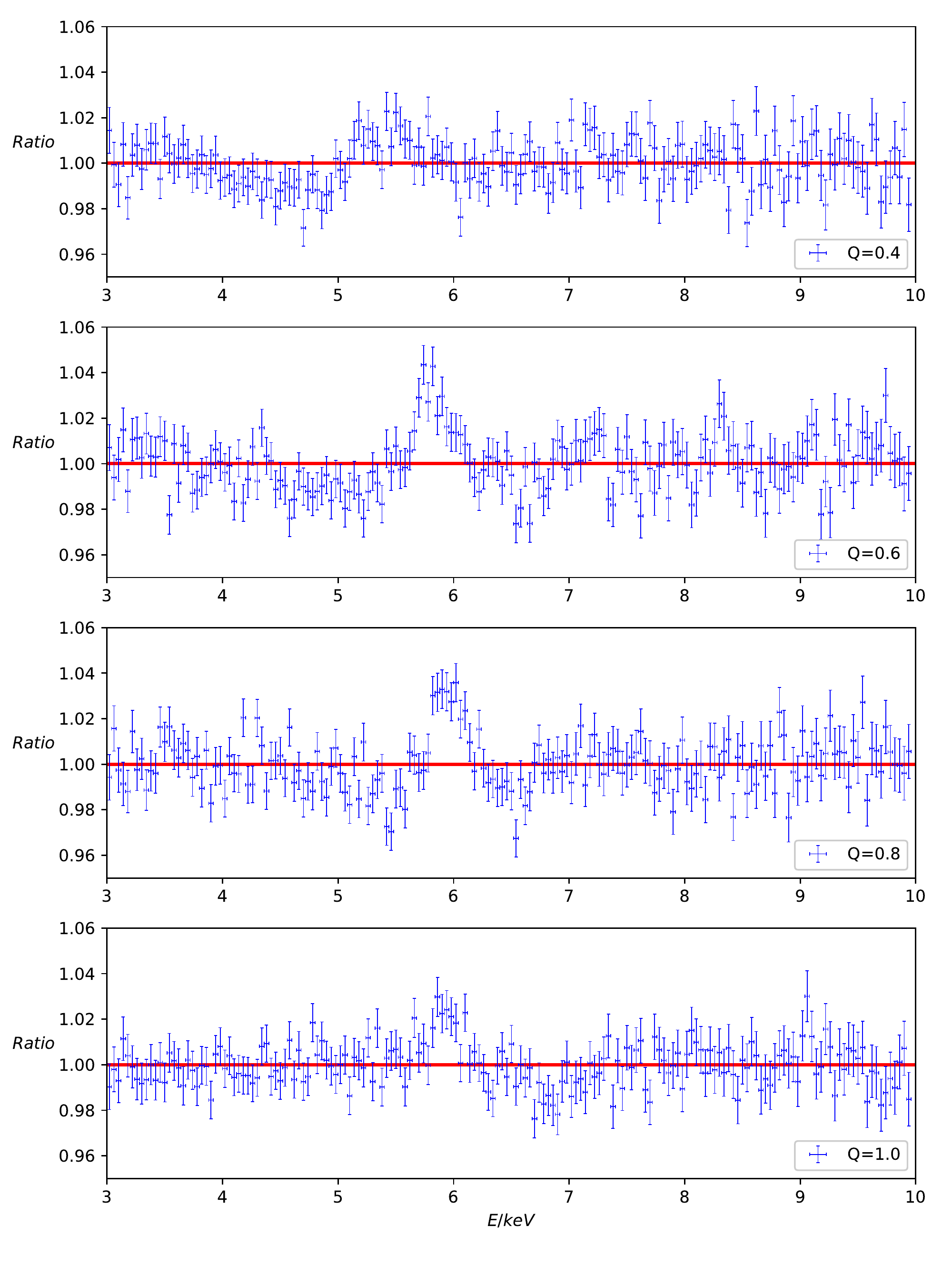}
\caption{\label{fig: rat-ultra} Ratio between data and the best-fit model for simulations 10--19 in the ultraspinning regime,~i.e.~$\beta>1$ and $a_{*}>1$, of VTG. The top-left set is for simulations 10--12 with spin parameter $a_{*}=2.0$ and charge $Q=0.4$, the top-right set is for simulations 13--15 with spin parameter $a_{*}=4.0$ and charge $Q=0.4$, and the bottom set is for simulations 16--19 with spin parameter $a_{*}=2.0$ and coupling parameter $\beta=6.0$. See text and Table~\ref{tab: ultra} for more details.}
\end{figure*}

We fit the simulated data with a power-law component and a Kerr iron line generated by {\sc relline}~\cite{2010MNRAS.409.1534D}. In the fitting we have six free parameters: the photon index of the power-law component $\Gamma$, the normalization of the power-law component, the spin parameter of the BH $a_{*}$, the inclination angle $\iota$, the emissivity index $q$, and the normalization of the iron line.

The best-fit values and the associated reduced $\chi^{2}$ of our 19 simulations are reported in Tables~\ref{tab: weak}--\ref{tab: ultra-2}. The ratios between the data and the best-fit model are shown in Figs.~\ref{fig: rat-weak}--\ref{fig: rat-ultra}. In the case of the weak coupling regime,~i.e.~$0<\beta<1$, it is clear that we can find good fits,~i.e.~the iron lines of BHs in VTG can be well modeled by iron lines of a Kerr BH. This is not unexpected since in the weak coupling regime the deformation from the Kerr spacetime is small. For the strong coupling regime,~i.e.~$\beta>1$, there seems to be a trend that the fits become worse as the spin parameter $a_{*}$ approaches its maximal value of $1$ and the coupling parameter $\beta$ becomes larger, but it is difficult from our sample size to definitively make this conclusion. In general, the fits in this regime are fair to poor. The fits in the ultraspinning regime,~i.e.~$\beta>1$ and $a_{*}>1$, are generally bad meaning that iron lines in the Kerr spacetime cannot be used to model the iron lines in the VTG spacetime in this regime. Note that while the reduced $\chi^{2}$ for simulations 14, 15, and 19 are not far from 1, the uncertainty on the spin spans almost the entire possible range. This implies that the Kerr iron line cannot fit the VTG iron line well, as the spin cannot be recovered, but is an integral parameter in determining the iron line shape. The ratio between data and best-fit model for simulations 14, 15, and 19 also support this conclusion as it is clear there is significant difference between the simulation and best-fit, particularly around 6 keV. Overall, we can conclude that observations with \textsl{NuSTAR} of the reflection spectrum should be able to test VTG gravity in the ultraspinning regime. It may also be possible to test the strong coupling regime, although it must be noted that since we have used a simple model compared to the full reflection spectrum the complexity of the full model may make it more difficult to test VTG in the strong coupling regime than depicted in this work. Finally, from our results it is fairly clear that the deviations introduced in the weak coupling regime are not significant enough to be able to be tested by reflection spectrum observations.

\section{Concluding Remarks}
\label{sec: conc}

In this work we have studied whether VTG gravity can be tested and constrained using observations of the BH X-ray reflection spectrum with current X-ray missions. As a preliminary exploration, for our model we only included a power-law component to represent the emission from a hot corona and a relativistically broadened K$\alpha$ iron line as the most prominent component that would be present in a full reflection spectrum. The spacetime of the BH was modeled with the BH solution of VTG found in~\cite{Filippini:2017kov}. Simulations of our model using different values of parameters relevant in the BH metric were done and then fit using the same reflection model but with the Kerr metric describing the BH spacetime. 

Our results suggest that VTG gravity can indeed be tested and constrained with current X-ray reflection spectrum observations, but only in the ultraspinning and (possibly) the strong coupling regimes. In the former case the iron line shows significant deviations from the iron line in a Kerr background that would almost certainly translate to observable deviations in the full reflection spectrum. For the latter, the deviations are not as large and it is not completely clear if they would still be observable in the full reflection spectrum. In the weak coupling regime, we find that the Kerr iron line can fit the VTG iron line quite well, implying that it would not be possible to distinguish lines between the two spacetimes in that regime.

Given our results an obvious next step would be to perform an analysis to test and place constraints on VTG gravity using current observations of the X-ray reflection spectra of stellar-mass and supermassive black holes. Such a study would need to use a more complex model(s), such as those used in~\cite{Cao:2017kdq, Tripathi:2018bbu}.

\acknowledgements

This work was supported by the National Natural Science Foundation of China (Grant No.~U1531117) and Fudan University (Grant No.~IDH1512060).~C.B. also acknowledges support from the  Alexander von Humboldt
Foundation.

\bibliographystyle{apsrev}
\bibliography{biblio}
\end{document}